\documentclass[aps,prl,showpacs,twocolumn,lineno,groupedaddress]{revtex4}  % for review

\ifx\pdftexversion\undefined
  \usepackage[dvips]{graphics}
\else
\usepackage[pdftex]{graphics}
\fi
\usepackage{graphicx}
\usepackage{dcolumn}   % needed for some tables
\usepackage{bm}        % for math
\usepackage{amssymb}   % for math

\newcommand{\beq}{\begin{equation}}
\newcommand{\eeq}{\end{equation}}
\newcommand{\barr}{\begin{eqnarray}}
\newcommand{\earr}{\end{eqnarray}}

\def\rperp{R_{\perp}}

\def\bq{\begin{quote}}
\def\eq{\end{quote}}

\def\spose#1{\hbox to 0pt{#1\hss}}
\def\lsim{\mathrel{\spose{\lower 3pt\hbox{$\mathchar"218$}}
 \raise 2.0pt\hbox{$\mathchar"13C$}}}
\def\gsim{\mathrel{\spose{\lower 3pt\hbox{$\mathchar"218$}}
 \raise 2.0pt\hbox{$\mathchar"13E$}}}

\def\bsbar{${\overline{B}_s^0}$}
\def\bs{${B_s^0}$}
\def\bd{${B^0}$}
\def\bsdec{${B_s^0 \rightarrow J/\psi~\phi}$}

\def\bddec{${B^0 \rightarrow J/\psi K^*}$}

\def\kstdec{$K^* \rightarrow K^{\pm} \pi^{\mp}$}

\def\D0{D\O }

\def\GeVp{ {\ifmmode \;{{\mbox{\mathrm GeV}} / {\mbox\mathrm c}} \else
${{\mbox{\mathrm GeV}} / {\mbox\mathrm c}}$ \fi }}
\def\MeVp{ {\ifmmode \;{{\mbox{\mathrm MeV}} / {\mbox\mathrm c}} \else
${{\mbox{\mathrm MeV}} / {\mbox\mathrm c}}$ \fi }}
\def\MeV{ {\ifmmode \;{{\mbox{\mathrm MeV}} / {\mbox\mathrm c}^2} \else
${{\mbox{\mathrm MeV}} / {\mbox\mathrm c}^2}$ \fi }}
\def\GeV{ {\ifmmode \;{{\mbox{\mathrm GeV}} / {\mbox\mathrm c}^2} \else
${{\mbox{\mathrm GeV}} / {\mbox\mathrm c}^2}$ \fi }}

\providecommand{\tabularnewline}{\\}

\begin{document}
%\leftline {Version  of June 9, 2005,   {\bf Deadline for comments: 15 Jun 05}} 
%\leftline {Primary authors: Avdhesh Chandra, Shashi Dugad, Daria Zieminska}
%\rightline{Comments to {\tt d0-run2eb-027@fnal.gov}}

%\preprint{FERMILB-PUB-xx/xxx-E}
%\preprint{FERMILAB-PUB-05-324-E}

\hspace{5.2in}
\mbox{FERMILAB-PUB-05-324-E}

\title{Measurement of the Lifetime Difference in the $B_s^0$ System}

% LIST_OF_AUTHORS_R2.TEX                 7/13/05            
%
\author{                                                                      
%% names begin here                                                           
V.M.~Abazov,$^{35}$                                                           
B.~Abbott,$^{72}$                                                             
M.~Abolins,$^{63}$                                                            
B.S.~Acharya,$^{29}$                                                          
M.~Adams,$^{50}$                                                              
T.~Adams,$^{48}$                                                              
M.~Agelou,$^{18}$                                                             
J.-L.~Agram,$^{19}$                                                           
S.H.~Ahn,$^{31}$                                                              
M.~Ahsan,$^{57}$
% ...
\\                                                                            
\vskip 0.30cm                                                                 
\centerline{(D\O\ Collaboration)}                                             
\vskip 0.30cm                                                                 
}                                                                             
\affiliation{                                                                 
\centerline{$^{1}$Universidad de Buenos Aires, Buenos Aires, Argentina}       
\centerline{$^{2}$LAFEX, Centro Brasileiro de Pesquisas F{\'\i}sicas,         
                  Rio de Janeiro, Brazil}                                     
\centerline{$^{3}$Universidade do Estado do Rio de Janeiro,                   
                  Rio de Janeiro, Brazil}                                     
\centerline{$^{4}$Instituto de F\'{\i}sica Te\'orica, Universidade            
                  Estadual Paulista, S\~ao Paulo, Brazil}                     
\centerline{$^{5}$University of Alberta, Edmonton, Alberta, Canada,           
               Simon Fraser University, Burnaby, British Columbia, Canada,}   
\centerline{York University, Toronto, Ontario, Canada, and                    
         McGill University, Montreal, Quebec, Canada}                         
\centerline{$^{6}$Institute of High Energy Physics, Beijing,                  
                  People's Republic of China}                                 
\centerline{$^{7}$University of Science and Technology of China, Hefei,       
                  People's Republic of China}                                 
\centerline{$^{8}$Universidad de los Andes, Bogot\'{a}, Colombia}             
\centerline{$^{9}$Center for Particle Physics, Charles University,            
                  Prague, Czech Republic}                                     
\centerline{$^{10}$Czech Technical University, Prague, Czech Republic}        
\centerline{$^{11}$Center for Particle Physics, Institute of Physics,         
                   Academy of Sciences of the Czech Republic,                 
                   Prague, Czech Republic}                                    
\centerline{$^{12}$Universidad San Francisco de Quito, Quito, Ecuador}        
\centerline{$^{13}$Laboratoire de Physique Corpusculaire, IN2P3-CNRS,         
                  Universit\'e Blaise Pascal, Clermont-Ferrand, France}       
\centerline{$^{14}$Laboratoire de Physique Subatomique et de Cosmologie,      
                  IN2P3-CNRS, Universite de Grenoble 1, Grenoble, France}     
\centerline{$^{15}$CPPM, IN2P3-CNRS, Universit\'e de la M\'editerran\'ee,     
                  Marseille, France}                                          
\centerline{$^{16}$IN2P3-CNRS, Laboratoire de l'Acc\'el\'erateur              
                  Lin\'eaire, Orsay, France}                                  
\centerline{$^{17}$LPNHE, IN2P3-CNRS, Universit\'es Paris VI and VII,         
                  Paris, France}                                              
\centerline{$^{18}$DAPNIA/Service de Physique des Particules, CEA, Saclay,    
                  France}                                                     
\centerline{$^{19}$IReS, IN2P3-CNRS, Universit\'e Louis Pasteur, Strasbourg,  
                France, and Universit\'e de Haute Alsace, Mulhouse, France}   
\centerline{$^{20}$Institut de Physique Nucl\'eaire de Lyon, IN2P3-CNRS,      
                   Universit\'e Claude Bernard, Villeurbanne, France}         
\centerline{$^{21}$III. Physikalisches Institut A, RWTH Aachen,               
                   Aachen, Germany}                                           
\centerline{$^{22}$Physikalisches Institut, Universit{\"a}t Bonn,             
                  Bonn, Germany}                                              
\centerline{$^{23}$Physikalisches Institut, Universit{\"a}t Freiburg,         
                  Freiburg, Germany}                                          
\centerline{$^{24}$Institut f{\"u}r Physik, Universit{\"a}t Mainz,            
                  Mainz, Germany}                                             
\centerline{$^{25}$Ludwig-Maximilians-Universit{\"a}t M{\"u}nchen,            
                   M{\"u}nchen, Germany}                                      
\centerline{$^{26}$Fachbereich Physik, University of Wuppertal,               
                   Wuppertal, Germany}                                        
\centerline{$^{27}$Panjab University, Chandigarh, India}                      
\centerline{$^{28}$Delhi University, Delhi, India}                            
\centerline{$^{29}$Tata Institute of Fundamental Research, Mumbai, India}     
\centerline{$^{30}$University College Dublin, Dublin, Ireland}                
\centerline{$^{31}$Korea Detector Laboratory, Korea University,               
                   Seoul, Korea}                                              
\centerline{$^{32}$CINVESTAV, Mexico City, Mexico}                            
\centerline{$^{33}$FOM-Institute NIKHEF and University of                     
                  Amsterdam/NIKHEF, Amsterdam, The Netherlands}               
\centerline{$^{34}$Radboud University Nijmegen/NIKHEF, Nijmegen, The          
                  Netherlands}                                                
\centerline{$^{35}$Joint Institute for Nuclear Research, Dubna, Russia}       
\centerline{$^{36}$Institute for Theoretical and Experimental Physics,        
                  Moscow, Russia}                                             
\centerline{$^{37}$Moscow State University, Moscow, Russia}                   
\centerline{$^{38}$Institute for High Energy Physics, Protvino, Russia}       
\centerline{$^{39}$Petersburg Nuclear Physics Institute,                      
                   St. Petersburg, Russia}                                    
\centerline{$^{40}$Lund University, Lund, Sweden, Royal Institute of          
                   Technology and Stockholm University, Stockholm,            
                   Sweden, and}                                               
\centerline{Uppsala University, Uppsala, Sweden}                              
\centerline{$^{41}$Lancaster University, Lancaster, United Kingdom}           
\centerline{$^{42}$Imperial College, London, United Kingdom}                  
\centerline{$^{43}$University of Manchester, Manchester, United Kingdom}      
\centerline{$^{44}$University of Arizona, Tucson, Arizona 85721, USA}         
\centerline{$^{45}$Lawrence Berkeley National Laboratory and University of    
                  California, Berkeley, California 94720, USA}                
\centerline{$^{46}$California State University, Fresno, California 93740, USA}
\centerline{$^{47}$University of California, Riverside, California 92521, USA}
\centerline{$^{48}$Florida State University, Tallahassee, Florida 32306, USA} 
\centerline{$^{49}$Fermi National Accelerator Laboratory, Batavia,            
                   Illinois 60510, USA}                                       
\centerline{$^{50}$University of Illinois at Chicago, Chicago,                
                   Illinois 60607, USA}                                       
\centerline{$^{51}$Northern Illinois University, DeKalb, Illinois 60115, USA} 
\centerline{$^{52}$Northwestern University, Evanston, Illinois 60208, USA}    
\centerline{$^{53}$Indiana University, Bloomington, Indiana 47405, USA}       
\centerline{$^{54}$University of Notre Dame, Notre Dame, Indiana 46556, USA}  
\centerline{$^{55}$Iowa State University, Ames, Iowa 50011, USA}              
\centerline{$^{56}$University of Kansas, Lawrence, Kansas 66045, USA}         
\centerline{$^{57}$Kansas State University, Manhattan, Kansas 66506, USA}     
\centerline{$^{58}$Louisiana Tech University, Ruston, Louisiana 71272, USA}   
\centerline{$^{59}$University of Maryland, College Park, Maryland 20742, USA} 
\centerline{$^{60}$Boston University, Boston, Massachusetts 02215, USA}       
\centerline{$^{61}$Northeastern University, Boston, Massachusetts 02115, USA} 
\centerline{$^{62}$University of Michigan, Ann Arbor, Michigan 48109, USA}    
\centerline{$^{63}$Michigan State University, East Lansing, Michigan 48824,   
                   USA}                                                       
\centerline{$^{64}$University of Mississippi, University, Mississippi 38677,  
                   USA}                                                       
\centerline{$^{65}$University of Nebraska, Lincoln, Nebraska 68588, USA}      
\centerline{$^{66}$Princeton University, Princeton, New Jersey 08544, USA}    
\centerline{$^{67}$Columbia University, New York, New York 10027, USA}        
\centerline{$^{68}$University of Rochester, Rochester, New York 14627, USA}   
\centerline{$^{69}$State University of New York, Stony Brook,                 
                   New York 11794, USA}                                       
\centerline{$^{70}$Brookhaven National Laboratory, Upton, New York 11973, USA}
\centerline{$^{71}$Langston University, Langston, Oklahoma 73050, USA}        
\centerline{$^{72}$University of Oklahoma, Norman, Oklahoma 73019, USA}       
\centerline{$^{73}$Brown University, Providence, Rhode Island 02912, USA}     
\centerline{$^{74}$University of Texas, Arlington, Texas 76019, USA}          
\centerline{$^{75}$Southern Methodist University, Dallas, Texas 75275, USA}   
\centerline{$^{76}$Rice University, Houston, Texas 77005, USA}                
\centerline{$^{77}$University of Virginia, Charlottesville, Virginia 22901,   
                   USA}                                                       
\centerline{$^{78}$University of Washington, Seattle, Washington 98195, USA}  
}                                                                             
%end                                                                          

\date{July 19, 2005}

\begin{abstract}

We present a study of the  decay  \bsdec.
%From a simultaneous fit to the distributions in the mass, 
%proper decay length, and transversity, we obtain 
We obtain the CP-odd fraction in the final state  at time zero, 
  $\rperp = 0.16 \pm 0.10$ (stat) $\pm$ 0.02 (syst),
the average lifetime of the (\bs, \bsbar) system, 
$\overline \tau (B_s^0) =1.39^{+0.13}_{-0.16}$ (stat) $^{+0.01}_{-0.02}$ (syst) ps,
and the relative width difference between the heavy and light mass eigenstates,
$\Delta \Gamma /\overline \Gamma \equiv (\Gamma_L - \Gamma_H)/\overline \Gamma
=0.24^{+0.28}_{-0.38}$ (stat) $^{+0.03}_{-0.04}$ (syst).
With   the additional constraint from the world average of the 
$B_s^0$ lifetime measurements using  semileptonic decays, 
we find 
%the one standard deviation range  reduced to 
$\overline \tau (B_s^0)= 1.39 \pm 0.06$ ~ps and
$\Delta \Gamma /\overline \Gamma = 0.25^{+0.14}_{-0.15}$. 
For the ratio of the  $B_s^0$ and $B^0$ lifetimes
we obtain
$\frac{\overline \tau(B_s^0)}{\tau(B^0)} = 0.91\pm0.09$ (stat) $\pm$ 0.003 
(syst).
%The data sample corresponds to an integrated luminosity of 450 pb$^{-1}$
%accumulated with the D\O\ detector at the Tevatron.
\end{abstract}
\maketitle

Within the framework of the standard model (SM), the $B_s^0$ mesons
 are expected
to mix in such a way  that the mass and decay width differences between the 
heavy and light eigenstates, 
$\Delta M \equiv M_H - M_L$ and
$\Delta \Gamma \equiv \Gamma_L - \Gamma_H$, are sizeable. The
mixing phase $\delta \phi$ is small and
to a good approximation the two mass eigenstates correspond to CP
eigenstates.  
New phenomena may alter  $\delta \phi$, leading to
a reduction of the observed $\Delta \Gamma /\overline \Gamma$
compared to the SM prediction~\cite{DFN2001}.

The decay \bsdec, proceeding through the quark process
$b\rightarrow c \bar c s$, 
gives rise to both CP-even and CP-odd final states.
It is possible to
separate the two CP components of the decay  \bsdec,
and thus  to measure the
lifetime difference, through a simultaneous study of
the time evolution 
and angular distributions of the decay products of the 
$J/\psi$ and $\phi$ mesons.
Angular distribution of $B_s^0 \rightarrow J/\psi(\rightarrow \mu^+ \mu^-)$ $\phi(\rightarrow K^+ K^-)$
involve three angles. Current statistics are such that the use 
of all three angles characterizing the final state is not yet beneficial.
 We use a variable particularly sensitive to 
separating the CP states, the cosine of the transversity polar angle 
(called "tranversity"), as defined below.

The analysis of data collected with the  D\O\ detector~\cite{run2det} 
at the Fermilab Tevatron 
collider presented in this Letter is 
an extension of a recently published 
 study~\cite{pedro} done under the 
single $B_s^0$  lifetime hypothesis.   
We perform an unbinned maximum likelihood fit to the data,
including the $B_s^0$ candidate mass, lifetime, and  transversity, 
in the decay sequence \bsdec, $J/\psi \rightarrow \mu^+ \mu^-$,
$\phi \rightarrow K^+ K^-$. 
We extract three parameters characterizing the \bs\ system and its
decay: $\overline \tau (B_s^0) =1/\overline \Gamma$, where
$\overline{\Gamma}\equiv(\Gamma_H+\Gamma_L)/2$;
$\Delta \Gamma/\overline \Gamma$;
% where $\Delta \Gamma \equiv \Gamma_L - \Gamma_H$; 
and $\rperp$, the relative rate of the
decay to the CP-odd states at time zero. 
%The subscripts $H$ and $L$ refer to  the $Heavy$ and $Light$ components
%of the $B_s^0$ system.
The {\em average} lifetimes of $B^0_s$ and $B^0$, as defined above,
are expected to be equal to within 1\%~\cite{tauratio}, and
their ratio is  determined by measuring the lifetime of $B^0$ in
the similar decay topology of $J/\psi K^*$.

%The D\O\ detector  comprises a 
%central-tracking system, a preshower detector, a calorimeter, 
%and a muon spectrometer~\cite{run2det}.
%The central-tracking system consists of a silicon microstrip tracker
%and a central fiber tracker, both located within a 2~T
%superconducting solenoidal magnet.  
%%Particle energies are measured in three liquid-argon/uranium calorimeters. 
%The muon spectrometer is outside
%the calorimetry and consists of a
%layer of tracking detectors and scintillation trigger counters
%inside of  1.8~T iron toroid magnets, followed by two similar layers outside
%the toroids. 
%It  covers the pseudorapidity range  $|\eta|<2$,
%where  $\eta = -\ln[\tan(\Theta/2)]$, and  $\Theta$ is the 
%polar angle with respect to the proton beam direction.

The data were collected between June 2002 and August 2004.
The  sample is selected  by requiring 
two reconstructed muons with a transverse momentum
$p_T>$ 1.5 GeV. Each muon is required to be detected  as a track segment 
in at least one layer of the muon system and to be matched to a central track.
One muon is required to  have segments both inside and outside the toroid.
%We reject runs where the muon or central tracking information was corrupted.
We require the events to satisfy a muon trigger that does not include any 
cuts
on the impact parameter.
The sample corresponds to an integrated luminosity of approximately 450 pb$^{-1}$.

To select the $B_{s}^0$ candidate sample, we apply the following kinematic
and quality cuts.
Minimum values of  momenta in the transverse plane for $B_{s}^0$, 
$\phi$, and $K$ mesons are set at
6.0 GeV,    1.5 GeV, and 0.7 GeV, respectively.
$J/\psi$ candidates are accepted if the invariant mass 
is in the range 2.90 -- 3.25 GeV.
% For events in the central 
%rapidity region (an event is considered to be central if the higher $p_T$
%muon has $|\eta_{\mu 1}|<1$), 
%we require the
%transverse momentum of the $J/\psi$ meson to exceed 4 GeV. 
Successful candidates are constrained to the average reconstructed
$J/\psi$ mass of 3.072 GeV.
Decay products of the $\phi$ candidates are required 
to  satisfy a  fit to a common vertex and to have an  invariant mass 
in the 
range 1.01 -- 1.03 GeV.
We require the ($J/\psi,\phi$) 
pair to be consistent with coming from a common vertex, and to have 
an invariant mass in the range 5.0 -- 5.8 GeV. 
In case of multiple $\phi$ meson 
candidates, we
select the one with the highest transverse momentum.
Monte Carlo (MC) studies show that the $p_T$ spectrum of the $\phi$ mesons
coming from $B_s^0$ decay is harder than the spectrum of a pair
of random tracks from the underlying event.  
We define the signed  decay length of a \bs\ meson $L^B_{xy}$ 
as the vector pointing
from the primary vertex to the decay vertex projected on the
\bs\ transverse momentum.
To reconstruct the primary vertex, 
we select   tracks with   $p_{T}>$ 0.3 GeV
that are not used as decay products of the $B_s$ candidate
and apply a constraint to the average beam spot position.
The  proper decay length, $ct$, is  
defined by the relation $ct = L^B_{xy}\cdot M_{B^0_s}/p_T$
where $M_{B^0_s}$ is the world average mass of the \bs\ meson~\cite{PDG}. 
The  distribution of the proper decay length  uncertainty $\sigma(ct)$  of \bs\ mesons 
 peaks around 25 $\mu$m. 
We accept events with $\sigma(ct) <60$  $\mu$m.
%Finally, we reject an event if the number of tracks other than muons
%in a cone $\Delta R$ around $J/\psi$ is  greater than 25.
There are 9699 events satisfying the above cuts.

The resulting invariant mass distribution of the ($J/\psi,\phi$) system is
shown in Fig.~\ref{fig:mass}. The curves are   
projections of the maximum likelihood fit, described below.
The fit assigns  513$\pm$33 events to  \bs\ decay.

Using the same data sample and analogous kinematic and quality cuts,
we select a sample of 1913 events of the decay sequence \bddec,  
$J/\psi \rightarrow \mu^+ \mu^-$, \kstdec,
and update the \bd\ lifetime  measurement
reported in Ref.~{\cite{pedro} with larger statistics. 

% The selection cuts are detailed in Table~\ref{cuts}.

%
\begin{figure}
\begin{center}\includegraphics[%
  width=8.0cm,
  height=8.5cm,
  keepaspectratio,
  trim=0 40 0 0
  ]{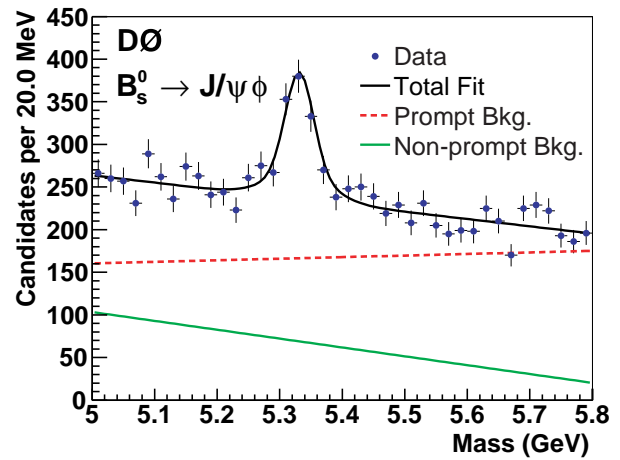}\end{center}

\caption{\label{fig:mass}
The invariant mass distribution of the ($J/\psi,\phi$) system for 
 \bs\ candidates. The curves are projections of the maximum likelihood fit (see text). 
}
\end{figure}

We perform a simultaneous
 unbinned maximum likelihood fit to the proper decay length, 
transversity, and mass. 
The likelihood function ${\cal L}$ is given by:
\begin{eqnarray}\label{eq:likelihood}
{\cal L} & = & \prod^{N}_{i=1}[ f_{\text {sig}}{\cal F}^i_{\text {sig}} + 
(1-f_{\text {sig}}){\cal F}^i_{\text {bck}}],
\end{eqnarray}
where $N$ is the total number of events, 
 ${\cal F}^i_{\text {sig}}$ (${\cal F}^i_{\text {bck}}$)
is the product of the mass, proper decay length, 
and the  transversity probability density functions for the signal
(background),  and $f_{\text {sig}}$ is the fraction 
of signal in the sample. 
Background is divided into two categories, based on their origin
and lifetime characteristics. 
``Prompt'' background is due to directly 
produced $J/\psi$ mesons accompanied by random tracks arising from 
hadronization.  This background is distinguished from ``non-prompt'' 
background, where the $J/\psi$ meson is a product of a B-hadron decay 
while the tracks forming the $\phi$ candidate emanate from a multibody 
decay of the same B hadron or from the underlying event.
We allow for independent parameters
for the two background components in mass, lifetime, and transversity.
There are nineteen free parameters in the fit.

%\vspace{2mm}
%{\bf Signal parametrization}
%\vspace{2mm}

For the signal mass distribution, we use
a sum of two Gaussian 
functions with a fixed  ratio of 
widths and normalizations, obtained in a fit to the signal-dominated
subset satisfying $ct/\sigma (ct)>5$. 
%$N_2/N_1$ = 0.35:0.65  and $\sigma_2/\sigma_1$ = 2.2, respectively.
%obtained by fitting the signal-dominated non-prompt subsample.
We  allow for two free parameters, the common mean value 
and the width of the narrow component.
The lifetime distribution of the signal is parametrized  by an
exponential convoluted with a Gaussian function with the width
taken from  the  event-by-event 
estimate of  $\sigma(ct)$.
To allow for the possibility of the lifetime uncertainty to be 
systematically
underestimated, we introduce a free scale factor.    

The transversity distribution of the signal is determined in 
the following way.
The time-dependent three-angle distribution 
 for the decay of {\it  untagged}  $B_s^0$ mesons, i.e., summed over
 $B^0_s$ and $\overline{B}^0_s$, 
expressed in terms of the linear polarization amplitudes
$|A_x(t)|$ and their relative phases $\delta_i$
is~\cite{ddf}:

\

\begin{widetext}
$$
\frac{d^3 \Gamma (t)}
{d \cos \theta~d \varphi~d \cos \psi}
\propto ~2 |A_0(0)|^2 e^{-\Gamma_L t}\cos^2
\psi
(1 - \sin^2 \theta\cos^2 \varphi)
\hfill{ } 
%$$
%$$
+ \sin^2 \psi \{ |A_\parallel(0)|^2  e^{-\Gamma_L t}
(1 - \sin^2 \theta \sin^2 \varphi)
$$
$$
+ |A_\perp(0)|^2 e^{-\Gamma_H t}\sin^2 \theta \} ~~~
%$$
%$$
+\frac{1}{\sqrt{2}}  \sin 2 \psi |A_0(0)|| A_\parallel(0)|
\cos(\delta_2-\delta_1)e^{-\Gamma_L t}
\sin^2
\theta  \sin 2 \varphi
$$
\beq\label{tripleuntagged}
%-|A_\parallel(0)|| A_\perp(0)|\cos\delta_1\sin^2\psi
%\sin2\theta\sin\varphi\biggr\}\frac{1}{2}\left(e^{-\Gamma_H t}-
%e^{-\Gamma_L t}\right)\delta\phi~~.
+\biggl\{\frac{1}{\sqrt{2}}|A_0(0)|| A_\perp(0)|\cos\delta_2\sin2\psi
\sin2\theta\cos\varphi
-|A_\parallel(0)|| A_\perp(0)|\cos\delta_1\sin^2\psi
\sin2\theta\sin\varphi\biggr\}\frac{1}{2}\left(e^{-\Gamma_H t}-
e^{-\Gamma_L t}\right)\delta\phi~.
\eeq
\end{widetext}
In the coordinate system of the $J/\psi$ rest frame 
 (where the $\phi$ meson moves in the $x$ direction,
 the $z$  axis is perpendicular to 
the decay plane of $\phi \to K^+ K^-$, and $p_y(K^+)\geq 0$),
the transversity polar and azimuthal angles 
$(\theta, \varphi)$ describe the
direction of the $\mu^+$, and $\psi$ is 
the angle between   $\vec p(K^+)$ and  $-\vec{p}(J/\psi)$ 
 in the $\phi$ rest frame.
%The quantity $\delta \phi$ is a CP-violating weak phase,
%due to the interference effects between 
%$B_s^0 - \overline B_s^0$ mixing and decay processes.
%In the standard model, $\delta \phi$ is negligibly small ($\approx -0.03).
%justifying the small-angle approximation in the above equation.

We model the acceptance in the three angles by
polynomials, with parameters determined using Monte Carlo simulations.
We have used  the {\sl SVV\_HELAMP} model
in the {\sc {EvtGen}} generator \cite{evtgen}, interfaced to the {\sc Pythia}
 program \cite{pythia}. Reconstructed events 
were reweighted to match the kinematic distributions with
the data.

To obtain the one-angle (transversity) distribution, we integrate 
the three-angle distribution  over the angles $\psi$ and $\varphi$. 
The resulting distribution
depends on one free parameter, $\rperp =|A_\perp(0)|^2 $. There is a small correction
term due to the nonuniformity of the acceptance in the angle $\varphi$,
which is proportional to $|A_0(0)|^2 - |A_\parallel(0)|^2$.
We use the CDF Collaboration measurement~\cite{CDF_run2} of this difference, 
0.355$\pm$0.066.

%\vspace{2mm}
%{\bf Background  parametrization}
%\vspace{2mm}

The lifetime shape  of the  background is described as a sum of
a prompt component, simulated as 
a Gaussian function centered at zero, and a non-prompt component,
simulated as a superposition of one exponential for the negative $ct$ 
region and two exponentials
for the positive $ct$ region, with free slopes and normalization.
The mass distributions of the backgrounds are parametrized by  first-order
polynomials. 
 The transversity distributions of backgrounds are parametrized 
as $(1+a_2\cos^2\theta + a_4\cos^4\theta)$.

%================
%\section{\label{sec:results}Results}
%================

\begin{figure}[htb]
\begin{center}\includegraphics[%
  width=8.0cm,
  height=8.5cm,
  keepaspectratio,
  trim=0 40 0 0
  ]{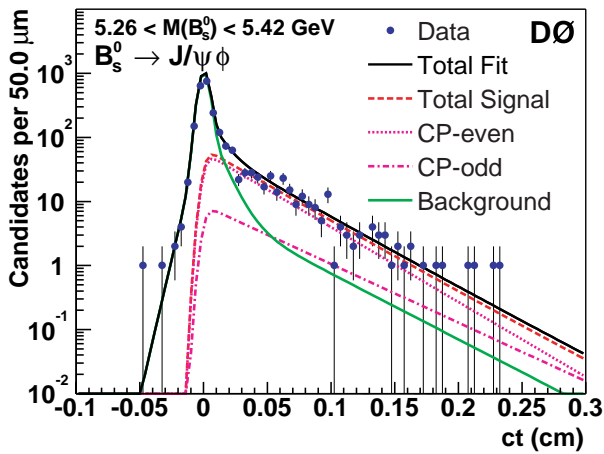}\end{center}

\caption{\label{fig:bs_lifetime_sig}
The proper decay length $ct$  of the \bs\ candidates
in the signal mass region.
The curves show  the signal contribution,  dashed (red); 
the background, lower solid line (green); and total, upper solid line (black)
in the signal  mass region.
}
\end{figure}

\begin{figure}[htb]
\begin{center}\includegraphics[%
  width=8.0cm,
  height=8.5cm,
  keepaspectratio,
  trim=0 40 0 0
  ]{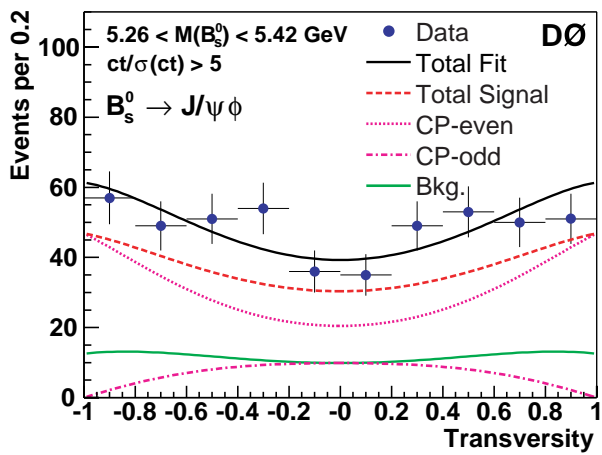}\end{center}

\caption{\label{fig:bs_trans_sig2}
The distribution of the cosine
of the transversity angle in the signal mass region, for ``non-prompt''
events, with the results of the maximum likelihood fit overlaid.
}
\end{figure}

\begin{figure}
\begin{center}\includegraphics[%
  width=8.0cm,
  height=8.5cm,
  keepaspectratio,
  trim=0 40 0 0
  ]{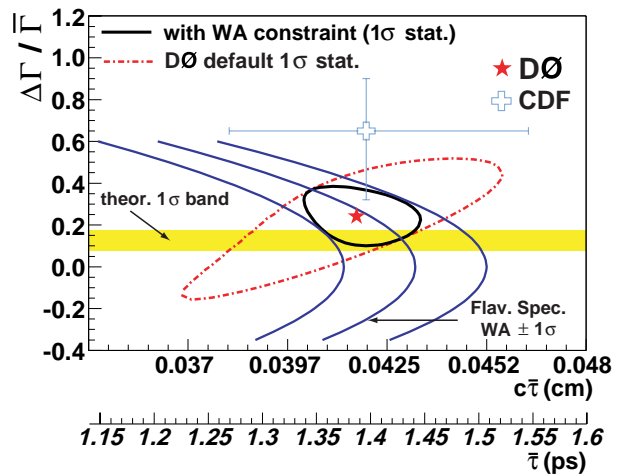}\end{center}

\caption{\label{fig:contour}
The  D\O\ default one-$\sigma$ (stat.) contour ($\delta\ln(L) = 0.5$) 
  compared to  a one-$\sigma$ 
band for the world average 
(WA) measurement based on flavor-specific decays,
$\tau_{fs} = 1.442 \pm 0.066$ ps. A simultaneous fit to our data 
and the WA
gives a one-$\sigma$ range
$\overline \tau (B_s^0)= 1.39 \pm 0.06$ ps
and  $\Delta \Gamma /\overline \Gamma =0.25^{+0.14}_{-0.15}$.
The SM theoretical prediction is shown as the horizontal band.
}
\end{figure}

%Results of this analysis are listed in Table~\ref{tab:results}.
Results of the fit are presented in 
Figs.~\ref{fig:mass} --
\ref{fig:contour}.
The proper decay length  distribution, 
and the transversity distribution,  
both with the fit results overlaid are shown 
in Figs.~\ref{fig:bs_lifetime_sig} and \ref{fig:bs_trans_sig2}. 
Figure~\ref{fig:contour} shows  the one standard deviation ( one-$\sigma$) 
contour for 
 $c \overline \tau (B_s^0)$ versus  $\Delta \Gamma/\overline \Gamma $.
It provides the best display of the uncertainty range for these 
correlated parameters. 
Our best fit returns $\rperp$= 0.16$\pm$0.10  and
  $\Delta \Gamma /\overline \Gamma = 0.24^{+0.28}_{-0.38}$ at
$\overline \tau(B^0_s)   =1.39^{+0.13}_{-0.16}$ ps.
%The most likely value of $\Delta \Gamma /\overline \Gamma$ rises with
%$c \overline \tau(B^0_s)$, keeping  the most likely value of the lifetime
%of the dominant light component stable within 2\% at  $\tau_L=1.24$ ps.

We  do a series of alternative fits, at 
 discrete values of $ \overline \tau(B^0_s)$. The results for 
$\Delta \Gamma /\overline \Gamma$,  its one standard deviation
range, and the corresponding value of the likelihood, are listed
in Table~\ref{tab:resultsfixedt}.  
We verify the  procedure by performing  fits on a sample of 
approximately 50,000 MC events passed through the full chain of  detector 
simulation, event reconstruction, and maximum likelihood fitting.
We see no bias in the event reconstruction or in the fitting procedure.
The fits reproduce the inputs ($c\tau=439$ $\mu$m, 
 $\Delta \Gamma /\overline \Gamma =0$, and a range of  $\rperp$
between 0 and 1)
correctly within the statistical precision
of $2\;\mu$m for
 $c\tau$, 0.01 for  $\rperp$,
and 0.025 for  $\Delta \Gamma /\overline \Gamma$.
We  test the sensitivity of the results to the parametrization of the
signal and background mass distributions  by varying the parameters of
the two-Gaussian  function.
To test the sensitivity of the results to the background model,
we  add  a quadratic term in the background mass distribution.
We find a non-negligible effect from the extra term in the non-prompt 
background
on  $c\tau$ and    $\Delta \Gamma /\overline \Gamma$.
We have also tested the sensitivity of the results to the assumption that the 
lifetime and the transversity distributions of background are independent
of mass. The effect of the uncertainty of  the detector alignment on the lifetime
measurement was estimated in Ref.~\cite{pedro}. 
The effects of systematic uncertainties are listed in Table \ref{syst}.

We  also conduct a test with an ensemble of
1000 pseudo-experiments with similar
statistical sensitivity, chosen from the distribution described by 
 Eq.~\ref{eq:likelihood},
with the same parameters
as obtained in this analysis.
Both the spread of uncertainties and
of the central values of the fit parameters are in good agreement 
with the  results
reported here.
% (see Fig.~\ref{fig:tmc_rperp}). 
Our results are consistent with the CDF Collaboration results~~\cite{CDF_run2},
also shown in Fig.~\ref{fig:contour}. 
%Our central value for $\Delta \Gamma /\overline \Gamma$ is lower by one
%standard deviation. 
%About 5\% of pseudo-experiments in our ensemble test 
% return  $\Delta \Gamma /\overline \Gamma >0.65$,
%i.e. above the CDF value~\cite{CDF_run2}. 
%If we constrain $\rperp$ to
%0.125 (CDF value of $|A_\perp (0)|^2$), we obtain  
%$\Delta \Gamma /\overline \Gamma = 0.32 \pm 0.22$ and 
% $\overline \tau (B_s^0)= 1.42\pm $0.09 ps.

\begin{table}

\caption{\label{tab:resultsfixedt}
Fit results for $\Delta \Gamma/\overline \Gamma$ at fixed values of 
$\overline \tau(B_s^0)$. For each assumed value of $\overline \tau(B_s^0)$,
 the likelihood as  function of
$\Delta \Gamma/\overline \Gamma$   is symmetric and parabolic.}
%\begin{ruledtabular}
%\begin{tabular}{cccccc}
\begin{tabular}{c r@{$\,\pm \,$}lc}

\hline
$\overline \tau(B_s^0)$ (ps) & \multicolumn{2}{c}  {$\Delta \Gamma/\overline \Gamma$} & $\Delta \ln(\cal L)$   \\
\hline

1.23             &  $-0.13$  & 0.15 & 0.51     \tabularnewline
%1.25             &  $-0.08$ & 0.16 & 0.4     \tabularnewline
1.27             &  $-0.03$  &0.17 & 0.32     \tabularnewline
%1.29             &  0.015  & 0.18 & 0.24     \tabularnewline
1.31           &  $ 0.07$ & 0.19 & 0.17     \tabularnewline
%1.32           &  0.09 & 0.20 & 0.13     \tabularnewline
%1.33           &  0.12 & 0.20 & 0.10     \tabularnewline
%1.34           &  0.14 & 0.21 & 0.07     \tabularnewline
1.35           &  ~0.16 & 0.21 & 0.04     \tabularnewline
%1.36           &  0.19 & 0.21 & 0.02     \tabularnewline
%1.37           &  0.21 & 0.21 & 0.01     \tabularnewline
%1.38           &  0.23 & 0.21 & 0.002     \tabularnewline
1.39           &  ~0.24 & 0.20 & 0.0     \tabularnewline
%1.40           &  0.26 & 0.19 & 0.005     \tabularnewline
%1.41           &  0.28 & 0.19 & 0.016     \tabularnewline
%1.42           &  0.30 & 0.19 & 0.033     \tabularnewline
1.43           &  ~0.31 & 0.19 & 0.06     \tabularnewline
%1.45           &  0.34 & 0.18 & 0.12     \tabularnewline
1.47           &  ~0.37 & 0.18 & 0.20     \tabularnewline
%1.50           &  0.42 & 0.18 & 0.36     \tabularnewline
1.51           &  ~0.43 & 0.18 & 0.42     \tabularnewline
1.55           &  ~0.48 & 0.18 & 0.69     \tabularnewline
%1.60           &  0.55 & 0.17 & 1.06     \tabularnewline
\hline

\end{tabular}
%\end{ruledtabular}
%\label{table:resultsfixedt}
\end{table}

\begin{table}[h!tb]
\caption {Sources of  systematic uncertainty. 
The numbers reflect  the variation of the fitted central values
associated with the one-$\sigma$ variation of the corresponding external input
parameters.
The second item includes contributions from the variation of the acceptance as a function
of $\varphi$ and $\psi$, as well as from a one-$\sigma$ variation 
of the quantity
$|A_0(0)|^2 - |A_\parallel(0)|^2$.
}
%The total effect of the systematic
%uncertainties quoted in the text is obtained by reading out the extrema of the
%curve encompassing all the alternative one-$\sigma$ contours corresponding 
%to the sources listed above.}
%\begin{ruledtabular}
\begin{tabular}{cccccc}
\hline
Source    &  $c\tau(B_s^0)$, $\mu$m & $\Delta \Gamma /\overline \Gamma$ & $\rperp$ \tabularnewline
\hline
Acceptance vs. $\cos \theta $         &$\pm$0.6  & $\pm$0.001 &$\pm$0.005\tabularnewline
Integration over $\varphi$, $\psi$    &$\pm$0.2  & $\pm$0.001 &$\pm$0.02   \tabularnewline
Procedure test                        &$\pm$2.0    &$\pm$0.025  &$\pm$0.01  \tabularnewline
Momentum scale                      &$-3.0$    &--  & --  \tabularnewline
Signal mass model                   &$\pm$1.0      &+0.009,$-0.017$     &$\pm$0.007 \tabularnewline
Background mass                &$-3.5$      &$+0.02$       &$-0.002$ \tabularnewline
Detector alignment                  &$\pm$2.0    & --    & -- \tabularnewline
Background  model            &$\pm$0.5  &$\pm$0.016  &$\pm$0.005\tabularnewline
\hline
Total                               &$-5.6$,$+3.1$  &$-0.04$, $+0.03$ &$\pm$0.02\tabularnewline
\hline
\end{tabular}
%\end{ruledtabular}

\label{syst}
\end{table}

 $B_s^0$  lifetime measurements from semileptonic (flavor-specific) data
provide an independent constraint on the average lifetime and lifetime
difference in the $B_s^0$  system. 
The world average~\cite{PDG} $B_s^0$ lifetime
%including the new, preliminary D\O\ measurement~\cite{d0sl},
%427$\pm$13$\pm$8 $\mu$m,  
is $\tau_{fs}=1/\Gamma_{fs} = 1.442 \pm 0.066$ ps.
This result is based   on  single-exponential fits  
in the flavor-specific decay channels, which determine the following 
relation~\cite{TeVbook} 
(shown in Fig. \ref{fig:contour})
of $\overline \Gamma$ and $\Delta \Gamma /\overline \Gamma$:
$\Gamma_{fs}=\overline \Gamma - (\Delta \overline \Gamma)^2/2\overline \Gamma + {\cal O}{(\Delta
 \overline \Gamma)^3/\overline \Gamma^2} \label{gammafs}$.
Applying  the above constraint to our measurement, we 
obtain
   $\overline \tau (B_s^0)= 1.39 \pm 0.06$~ps
%($c\overline \tau (B_s^0) = 418^{+17}_{-19}$ $\mu$m),
 and 
$\Delta \Gamma /\overline \Gamma = 0.25 ^{+0.14}_{-0.15}$.
This result is consistent with the SM expectation~\cite{lenz}
of 0.12 $\pm$ 0.05.

All the results presented above are obtained under 
a tacit assumption that the CP-violating phase
is negligible, as predicted by the SM
($\delta \phi =\phi_{\mbox{{\scriptsize CKM}}}=-0.03$). 
Future improvements on the  measurement of $\Delta \Gamma /\overline \Gamma$ 
 may exclude models
predicting  large deviations of $\delta \phi$ from the SM value.  

In summary, we have measured 
the CP-odd fraction for the decay \bsdec,  and
 the correlated parameters of the average lifetime 
of the (\bs, \bsbar) system   $\overline \tau(B^0_s) = 1/\overline \Gamma$,  
and the relative  width difference $\Delta \Gamma /\overline \Gamma$,
or, equivalently, the mean lifetimes of the light and heavy $B_s^0$
eigenstates, 
%which in the SM coincide with CP-even and CP-odd states,
respectively. We obtain

\vspace{0.1cm}

\centerline {$\rperp = 0.16$ $\pm$ 0.10 (stat) $\pm$ 0.02 (syst),}

%\vspace{0.2cm}

\centerline {$\Delta \Gamma /\overline \Gamma =0.24^{+0.28}_{-0.38}$ (stat)
$^{+0.03}_{-0.04}$ (syst),}

%\vspace{0.2cm}

\centerline {$\overline \tau(B^0_s)   =1.39^{+0.13}_{-0.16}$ (stat)
$^{+0.01}_{-0.02}$ (syst) ps,}

%\vspace{0.2cm}

\centerline {$\tau_L =1.24^{+0.14}_{-0.11}$ (stat)
$^{+0.01}_{-0.02}$ (syst) ps,}

%\vspace{0.2cm}
\centerline {$\tau_H =1.58^{+0.39}_{-0.42}$ (stat)
$^{+0.01}_{-0.02}$ (syst) ps.} 

%\vspace{0.2cm}

\vspace{0.1cm}

We have updated the  measurement of  the mean  lifetime 
of the $B^0$  meson with doubled statistics.
With the systematic uncertainty estimated in Ref.~\cite{pedro},
the updated measurement is

\vspace{0.1cm}

\centerline 
{$\tau (B^0)$ = 1.530 $\pm$ 0.043 (stat) $\pm$ 0.023 (syst) ps.}

\vspace{0.1cm}
For the ratio of the average $B_s^0$ lifetime to the $B^0$ lifetime,
we obtain

%\vspace{0.1cm}
\centerline 
{$\frac{\overline \tau(B_s^0)}{\tau(B^0)}$ = 0.91 $\pm$ 0.09 (stat) $\pm$ 0.003 (syst).}
\vspace{0.1cm}

%\vspace{0.1cm}
%
%Using our results for $\Delta \Gamma /\overline \Gamma$ and 
%$\overline \tau(B_s^0)$, and applying a constraint on this pair of 
%parameters from the existing semileptonic 
%(i.e. flavor-specific) measurements
%~\cite{PDG}, 
%we obtain:
%
%\vspace{0.1cm}
%
%
%\centerline
%{$   \overline \tau = 1.39^{+0.06}_{-0.06}$   ~ps,}
%
%
%\vspace{0.2cm}
%
%\centerline
%{$\Delta \Gamma /\overline \Gamma= 0.25 ^{+0.14}_{-0.15}$.}
%
%
%\vspace{0.1cm}
%
%
%Comparison of our results with the lifetime measurements from semileptonic
%decays, and with the theoretical prediction for $\Delta \Gamma _{\mbox{{\scriptsize CP}}}$
%leads to the following  best estimate for the cosine of the mixing phase,
%
%\vspace{0.2cm}
%%\beq
%\centerline
%{$|\cos(\delta \phi)| = 1.52^{+0.73}_{-0.65}.$}
%%\eeq
\vspace{0.2cm}

% acknowledgement_paragraph_r2.tex                5/17/05
%
We thank the staffs at Fermilab and collaborating institutions, 
and acknowledge support from the 
DOE and NSF (USA);
CEA and CNRS/IN2P3 (France);
FASI, Rosatom and RFBR (Russia);
CAPES, CNPq, FAPERJ, FAPESP and FUNDUNESP (Brazil);
DAE and DST (India);
Colciencias (Colombia);
CONACyT (Mexico);
KRF (Korea);
CONICET and UBACyT (Argentina);
FOM (The Netherlands);
PPARC (United Kingdom);
MSMT (Czech Republic);
CRC Program, CFI, NSERC and WestGrid Project (Canada);
BMBF and DFG (Germany);
SFI (Ireland);
Research Corporation,
Alexander von Humboldt Foundation,
and the Marie Curie Program.
%
   % input acknowledgement

%\input bs_appendix

\end{document}